\documentclass[prd,aps,twocolumn,showpacs,nofootinbib,floatfix]{revtex4-1}

\usepackage{amsfonts}
\usepackage[colorlinks=true,citecolor=blue, linkcolor=blue,urlcolor=blue]{hyperref}
\usepackage{color,graphicx,amsfonts,multirow}
\usepackage{amsmath}

\begin{document}

\title{Production of doubly heavy baryon at the Muon-Ion Collider}

\author{Xue-Yun Zhao$^{1}$}
\email{zhaoxy@stu.cqu.edu.cn}
\author{Lei Guo$^{1}$}
\email{guoleicqu@cqu.edu.cn}
\author{Xu-Chang Zheng$^{1}$}
\email{zhengxc@cqu.edu.cn}
\author{Huan-Yu Bi$^{2}$}
\email{bihy@pku.edu.cn}
\author{Xing-Gang Wu$^{1}$}
\email{wuxg@cqu.edu.cn}
\author{Qi-Wei Ke$^{1}$}
\email{keqw@cqu.edu.cn}

\affiliation{$^1$ Department of Physics, Chongqing University, Chongqing 401331, P.R. China}
\affiliation{$^2$ Center for Theoretical Physics $\&$ School of Physics and Optoelectronic Engineering, Hainan University, Haikou, 570228, China}


\begin{abstract}

This study forecasts the production of doubly heavy baryons, $\Xi_{cc}$, $\Xi_{bc}$, and $\Xi_{bb}$, within the nonrelativistic QCD framework at the Muon-Ion Collider (MuIC). It examines two production mechanisms: photon-gluon fusion ($\gamma + g \to (QQ')[n] +\bar{Q} +\bar{Q'}$) and extrinsic heavy quark channels ($\gamma + Q \to (QQ')[n] + \bar{Q'}$), where $Q$ and $Q'$ denote heavy quarks ($c$ or $b$) and $(QQ')[n]$ represents a diquark in specific spin-color configurations. The diquark fragments into $\Xi_{QQ'}$ baryons with high probability. For $\Xi_{cc}$ and $\Xi_{bb}$, the relevant configurations are $[^1S_0]_{\textbf{6}}$ (spin-singlet and color-sextuplet) and $[^3S_1]_{\bar{\textbf{3}}}$ (spin-triplet and color-antitriplet). For $\Xi_{bc}$, the configurations are $[^1S_0]_{\bar{\textbf{3}}}$, $[^1S_0]_{\textbf{6}}$, $[^3S_1]_{\bar{\textbf{3}}}$, and $[^3S_1]_{\textbf{6}}$. The study compares total and differential cross-sections for these channels, highlighting their uncertainties. The results indicate that the extrinsic heavy quark channel, particularly the $[^3S_1]_{\bar{\textbf{3}}}$ configuration, dominates $\Xi_{QQ'}$ production, though other diquark states also contribute significantly. Using quark masses $m_c = 1.80 \pm 0.10$ GeV and $m_b = 5.1 \pm 0.20$ GeV, the study estimates annual event yields at MuIC ($\sqrt{s} = 1$ TeV, luminosity ${\mathcal L}\simeq 40$ ${\rm fb}^{-1}$) of $(3.67^{+1.29}_{-0.91}) \times 10^9$ for $\Xi_{cc}$, $(2.24^{+0.28}_{-0.20}) \times 10^8$ for $\Xi_{bc}$, and $(3.00^{+0.64}_{-0.56}) \times 10^6$ for $\Xi_{bb}$. These findings suggest that MuIC will significantly enhance our understanding of doubly heavy baryons.
\end{abstract}

\maketitle

\section{Introduction}

Lepton-hadron deep inelastic scattering (DIS) is an essential technique for investigating the internal structure of nucleons and nuclei. Over the years, DIS experiments have uncovered the quark and gluon substructure and their momentum distribution within fast-moving nucleons. To further delve into the three-dimensional quark-gluon dynamics governed by quantum chromodynamics (QCD), a high-energy, high-luminosity polarized electron-ion collider (EIC) has been approved for construction at Brookhaven National Laboratory (BNL) by the late 2020s \cite{Aschenauer:2017jsk}, marking a top priority in U.S. nuclear physics. The EIC will facilitate polarized electron-proton and electron-nucleus collisions at center-of-mass energies up to 140 GeV \cite{Aschenauer:2017jsk, Accardi:2012qut}, establishing a new QCD frontier. This facility will address key questions about the origin of nucleon spin, mass, and QCD phenomena at high parton densities. Meanwhile, CERN's proposed Large Hadron-electron Collider (LHeC) \cite{LHeC:2020van} aims to explore TeV energy DIS with high luminosities, with the Future Circular Collider (FCC) set to include electron-hadron collisions at $\sqrt{s} = 3.5$ TeV \cite{FCC:2018byv}, utilizing the LHeC's electron beam.

The muon collider proposal has garnered renewed interest in the particle physics community in recent years due to its potential to achieve very high energies in a compact tunnel (e.g., the size of the LHC) at relatively low costs. The Muon-Ion Collider (MuIC) \cite{Acosta:2021qpx, Acosta:2022ejc} is a proposed project to be built at BNL, intended to succeed the EIC in the 2040s. MuIC aims to realize the next generation of lepton-hadron (ion) colliders at TeV scales based on existing hadron collider facilities. The BNL facility has the capability to support a muon storage beam with an energy of up to approximately 1 TeV using current magnet technology. When this beam collides with 275 GeV, the MuIC's center-of-mass energy of around 1 TeV will significantly expand the kinematic range of deep inelastic scattering physics at the EIC (with polarized beams) by more than an order of magnitude in $Q^{2}$ and $x$. This will open a new frontier in QCD, addressing numerous fundamental scientific questions in nuclear and particle physics. This range is comparable to that of the proposed LHeC at CERN, but with different lepton and hadron kinematics, ion species, and beam polarization. Furthermore, developing a MuIC at BNL will concentrate global R\&D efforts on muon collider technology and act as a prototype for a future muon-antimuon collider at energies of ${\cal O}(10)$ TeV. This is seen as a promising option to achieve the next high-energy frontier in particle physics at a more affordable cost and with a smaller footprint than a future circular hadron collider. In this article, we will investigate whether significant amounts of doubly heavy baryon events can be produced at the MuIC.

Doubly heavy baryons, which contain two heavy quarks, have a simplified structure akin to heavy quarkonia, making them suitable for detailed theoretical analysis. The SELEX Collaboration \cite{SELEX:2002wqn, SELEX:2004lln} first proposed the existence of $\Xi_{cc}^{+}$ in 2002 and 2005. More recently, in 2017, the LHCb Collaboration identified another doubly heavy baryon, $\Xi_{cc}^{+}$, through the decay mode $\Xi_{cc}^{++} \to \Lambda_{c}^{+} K^{-} \pi^{+} \pi^{+}$ \cite{Yu:2017zst}, with $\Lambda_{c}^{+} \to p K^{-} \pi^{+}$ \cite{LHCb:2017iph}. Further confirmation came from the LHCb Collaboration, which verified this baryon's existence via the decay channel $\Xi_{cc}^{++} \to \pi^{+}\Xi_{c}^{+}$ \cite{LHCb:2018pcs, LHCb:2018zpl}. These discoveries make doubly heavy baryons a crucial area for studying QCD. Due to strong interaction confinement, their production involves nonperturbative effects beyond perturbative QCD. In the work \cite{Ma:2003zk}, the nonrelativistic QCD (NRQCD) \cite{Bodwin:1994jh} factorization framework was employed to describe the production process. This approach separates the process into two stages: the perturbative generation of a heavy-quark pair in a specific quantum state, referred to as a diquark, and its subsequent nonperturbative transition into a baryon. By expanding in the small velocity $(v_{Q})$ of the heavy quark in the baryon's rest frame, two leading-order states of ($cc$)-diquarks are identified: $[^{3}S_{1}]_{\bar{\textbf{3}}}$ and $[^{1}S_{0}]_{\textbf{6}}$, each associated with a corresponding long-distance matrix element (LDME), namely $h_{\bar{\textbf{3}}}$ and $h_{\textbf{6}}$. $[^{3}S_{1}]_{\bar{3}}$($[^{1}S_{0}]{6}$) represents a (cc)-diquark in the S-wave $^{3}S_{1}$($^1S_{0}$) and in the ${\bar{\textbf{3}}}$(${\textbf{6}}$) color state, while $h_{\bar{\textbf{3}}}$($h_{\textbf{6}}$) depicts its nonperturbative transition probability into the baryon.

Extensive theoretical studies have explored the production of doubly heavy baryons \cite{Baranov:1995rc, Berezhnoy:1996an, Jiang:2012jt, Jiang:2013ej, Chen:2014frw, Yang:2014ita, Yang:2014tca, Martynenko:2013eoa, Zheng:2015ixa, Bi:2017nzv, Sun:2020mvl, Chen:2014hqa, Chen:2019ykv, Chen:2018koh, Martynenko:2014ola, Koshkarev:2016acq, Koshkarev:2016rci, Groote:2017szb, Berezhnoy:2018bde, Brodsky:2017ntu, Berezhnoy:2018krl, Wu:2019gta, Qin:2020zlg, Niu:2018ycb, Niu:2019xuq, Zhang:2022jst, Luo:2022jxq, Luo:2022lcj, Ma:2022cgt, Ma:2022ger, Niu:2023ojf, Li:2022mxp, Zhan:2023vwp, Zhan:2023jfm} through direct channels in $pp$, $ep$, $\gamma \gamma$, and $e^{+}e^{-}$ collisions, as well as indirectly via Higgs, $W$, $Z$ bosons, and top quark decays. The GENXICC \cite{Chang:2007pp, Chang:2009va, Wang:2012vj} generator has been developed to simulate hadroproduction in $pp$ collisions. Additionally, the muon-ion collider may also be a potential machine to probe the properties of doubly heavy baryons. The photoproduction mechanism dominates the production of $c/b$-quark at the MuIC, and the doubly heavy baryon can thus be primarily generated via the photoproduction channels $\gamma + g \to \Xi_{QQ'}+\bar{Q} +\bar{Q'}$ and $\gamma + Q \to \Xi_{QQ'}+\bar{Q'}$.

The photoproduction of $\Xi_{QQ'}$ can be divided into three steps. Using the $\gamma+g$ channel as an example, the first step involves producing $Q\bar{Q}$ and $Q'\bar{Q'}$ pairs, with the heavy quarks $Q$ and $Q'$ needing to be in the color and spin configuration $[n]$, The second step is the fusion of the $QQ'$ pair into a bound diquark $(QQ')[n]$ with a certain probability (The quark pairs in the color-sextuplet state experience a repulsive potential, making it theoretically impossible to form a binding color-sextuplet diquark. However, even though the quark pairs in the color-sextuplet state are mutually repulsive, they can still fragment into doubly heavy baryons.); the third step involves the diquark evolving into a doubly heavy baryon $\Xi_{QQ'}$ by capturing a light quark from the vacuum or by emitting/capturing an appropriate number of gluons. The first step can be calculated perturbatively, as the gluon should be hard enough to produce the heavy quark-antiquark pair. For the second step, the transition probability is described by a nonperturbative NRQCD matrix element. We use $h_{\textbf{6}}$ and $h_{\bar{\textbf{3}}}$ to denote the matrix elements for the production of a color-sextuplet (${\textbf{6}}$) and a color-antitriplet (${\bar{\textbf{3}}}$) diquark, respectively. Here, we do not differentiate between the matrix elements of the $^1S_0$ and $^3S_1$ states, as the spin-splitting effect is minimal \cite{Zhang:2011hi, Jiang:2012jt}. For the third step, it is typically assumed that the efficiency of the evolution from a $(QQ')[n]$ diquark to a doubly heavy baryon $\Xi_{QQ'}$ is $100\%$, a process referred to as "direct evolution". Reference \cite{Chen:2014frw} has examined both direct evolution and "evolution via fragmentation", which incorporates the fragmentation function. The authors concluded that direct evolution is highly accurate and sufficiently effective for studying the production of doubly heavy baryons. Consequently, we adopt the direct evolution approach in our calculations.

Given that the predicted production rate of $\Xi_{cc}$ is significantly lower than the SELEX measurements, the authors of Refs.\cite{Chang:2006eu, Chang:2006xp, Chang:2007pp} proposed considering both extrinsic and intrinsic charm production mechanisms to narrow the gap between theoretical and experimental results. It is noted that the intrinsic charm's contribution to the cross section of the $\gamma + c$ channel is less than $0.1\%$, even if the intrinsic $c$-component density in the proton is as high as $1\%$ \cite{Brodsky:1981se, Brodsky:1980pb}. Following the suggestions in Refs.\cite{Chang:2006eu, Chang:2007pp} and based on $B_c$ baryon photoproduction, we will focus on the channels $\gamma + g \to \Xi_{QQ'} +\bar{Q} +\bar{Q'}$ and $\gamma + Q \to \Xi_{QQ'}+\bar{Q'}$. In these channels, the intermediate diquark $(QQ')[n]$ can be $(cc/bb)_{\textbf{6}}[^1S_0]$, $(cc/bb)_{\bar{\textbf{3}}}[^3S_1]$, $(bc)_{{\bar{\textbf{3}}}/{\textbf{6}}}[^1S_0]$, or $(bc)_{{\bar{\textbf{3}}}/{\textbf{6}}}[^3S_1]$. Other diquark configurations, such as $(cc/bb)_{\textbf{6}}[^3S_1]$ and $(cc/bb)_{\bar{\textbf{3}}}[^1S_0]$, are prohibited due to Fermi-Dirac statistics for identical particles.

In our study, we focus on the leading-order (LO) contribution in the NRQCD framework. While higher-order $v^2$ corrections, including $1/m_c$ suppressed terms, could provide additional refinements, previous studies suggest that the LO terms capture the dominant physics. A full next-to-leading order (NLO) calculation, including power-suppressed terms, is beyond the scope of this work but remains an interesting direction for future studies.

The remainder of the paper is structured as follows: Section II details the calculation methodology. Section III provides numerical results, discusses theoretical uncertainties, and offers insights. Finally, Section IV presents a concise summary.

\section{Calculation technology}

\begin{widetext}
    \begin{eqnarray}
    &&d\sigma( \mu +P\to \Xi_{QQ'} + X) \nonumber \\
    =
    &&\sum_{[n]}
    \langle{\cal O}^{\langle QQ' \rangle}[n]\rangle
        \bigg \{    f_{\gamma/\mu}(x_1)  (N_{A}f^{h/A}_{g/P}(x_2,\mu_f))\otimes d\hat{\sigma}({\gamma+g}\to (QQ')[n] + X)  \nonumber \\
    &&~+ f_{\gamma/\mu}(x_1)  \Big[N_{A}\Big( f^{h/A}_{Q/P}(x_2, \mu_f) - f^{h/A}_{Q/P}(x_2, \mu_f)_{\rm SUB} \Big)\otimes
        d\hat{\sigma}({\gamma+Q}\to (QQ')[n] + X)+ Q\longleftrightarrow Q' \Big]\frac{1}{1+\delta_{QQ'}} \bigg \},~~~
    \label{eq1}
    \end{eqnarray}
\end{widetext}

In the photoproduction mechanism, the initial photon is emitted by the muon and can be described using the Weizsäcker-Williams approximation (WWA) \cite{vonWeizsacker:1934nji, Williams:1934ad, Frixione:1993yw}. When addressing the extrinsic heavy-quark mechanism, it is crucial to avoid ``double counting" between the $\gamma + g$ and extrinsic $\gamma + Q$ channels. An effective method for handling the extrinsic heavy quark is the application of the general-mass variable-flavor-number scheme (GM-VFNs) \cite{Olness:1987ep, Aivazis:1993kh, Aivazis:1993pi, Amundson:2000vg, Kniehl:2005mk}. According to the pQCD factorization theorem, the cross section for $\Xi_{QQ'}$ photoproduction within the GM-VFNs framework is expressed as Eq.(\ref{eq1})

where $f^{h/A}_{i/P}(x_2,\mu_f)$ represents the parton distribution function (PDF) of parton $i$ within a proton $P$, and $\mu_f$ is the factorization scale. The function $f_{\gamma/\mu}(x_1)$ denotes the photon density function, while $d\hat{\sigma}({\gamma+i}\to (QQ')[n] + X)$ is the hard cross section for the partonic process $\gamma+i\to (QQ')[n]+X$. The nonperturbative matrix element $\langle{\cal O}^{\langle QQ' \rangle}[n]\rangle$ represents the transition probability from the $(QQ')[n]$-quark pair to the desired baryon $\Xi_{QQ'}$. Here $\delta_{QQ'}=1 (0)$ when $Q=Q'$ ($Q\neq Q'$). Given that we employ the direct evolution scheme, the matrix elements $\langle{\cal O}^{\langle QQ' \rangle}[n]\rangle$ are either $h_{\bar{\textbf{3}}}$ or $h_{\textbf{6}}$, respectively.

The function $N_{A}f^{h/A}_{(g,Q)/P}(x_2,\mu_f)$ represents the effective parton distribution functions (PDFs) for the nucleus $\rm{A}$. It describes the parton density of a bound nucleon $h$ in nucleus $A$, carrying a fraction $x_2$ of the hadron momentum at the factorization scale $\mu_{f}$. Here, $h$ refers to the nucleon, proton, or neutron. $N_{A}$ denotes the atomic number of the incident nucleus. For example, $N_{Au} = 197$ for the gold nucleus ($^{197}_{79}Au$). Various PDFs models have been proposed to study heavy-ion collisions, including the Heavy-Ion Jet INteraction Generator (HIJING) model \cite{Wang:1991hta}, A Multiphase Transport (AMPT) model \cite{Lin:2004en}, the Monte Carlo Glauber Model \cite{Alver:2008aq, Broniowski:2007nz, Rybczynski:2013yba, Loizides:2017ack}, and others. Following the approach of the CTEQ group \cite{Kovarik:2015cma}, we adopt the PDFs of a bound nucleon in a nucleus as the heavy ion PDFs. In our calculation, we assume the same PDFs for protons and neutrons. This approximation is motivated by isospin symmetry, which relates the neutron PDFs to the proton PDFs via $f_u^p(x) = f_d^n(x)$, and the dominance of gluon distributions at high energy scales, where the difference between proton and neutron PDFs is negligible.

Under the condition of applying a small-angle cut to the scattered muon, the photon density function, characterized by the Weizsäcker-Williams approximation (WWA), is given by~\cite{Frixione:1993yw}
\begin{eqnarray}
f_{\gamma/\mu}(x)= \frac{\alpha}{2 \pi} \bigg [ \frac{1+(1-x)^2}{x} {\rm ln} \frac{Q^2_{\rm max}}{Q^2_{\rm min}} +\nonumber \\
 2 m_{\mu}^2 x \bigg (\frac{1}{Q^2_{\rm max}} -\frac{1}{Q^2_{\rm min}}\bigg) \bigg ],\label{wwaf}
\end{eqnarray}
where $x={E_{\gamma}}/{E_{\mu}}$, $E_{\gamma}$ and $E_{\mu}$ are photon and muon energies. $\alpha$ represents the fine structure constant, and $m_{\mu}$ denotes the muon mass, we do not neglect the higher-order terms of the muon mass. $Q^2_{\rm min}$ and $Q^2_{\rm max}$ are given by
\begin{align}
Q^2_{\rm min} &= \frac{m_{\mu}^2x^2}{1-x},\nonumber\\
Q^2_{\rm max} &= \frac{E_{\mu}(1+\beta)(A^2-m^{2}_{\mu})^2}{4A^3}\theta^{2}_{c}+Q^2_{\rm min},\label{theta}  
\end{align}
where $\beta=\sqrt{1-\frac{m^{2}_{\mu}}{E^{2}_{\mu}}}$, $A=E_{\mu}(1+\beta)(1-x)$, the scattering angle cut $\theta_c$ is determined by experiment \cite{Klasen:1997br, Klasen:2002xb}.

The subtraction term  $f_{Q/P}(x_2, \mu_f)_{\rm SUB}$ in Eq.(\ref{eq1}) is defined as
\begin{equation}
f_{Q/P}(x_2, \mu_f)_{\rm SUB}= \int_{x_2}^1 f_{g/P}(x_2/y, \mu_f) f_{Q/g}(y, \mu_f) \frac{dy}{y},
\end{equation}
where $f_{Q/g}(y, \mu_f)$ represents the $Q$-quark distribution function inside an on-shell gluon, which can be expanded perturbatively in $\alpha_s$. At the $\alpha_s$-order, $f_{Q/g}(y, \mu_f)$ is given by
\begin{eqnarray}
f_{Q/g}(y,\mu_f)=\frac{\alpha_s(\mu_f)}{2\pi} {\rm{ln}} \frac{\mu_f^2}{m_Q^2}P_{g\to Q}(y),
\end{eqnarray}
where $P_{g\to Q}(y)=\frac{1}{2}(1-2y+2y^2)$ is the $g \to Q \bar{Q}$ splitting function.

The hard partonic cross section is expressed as
\begin{equation}
d\hat{\sigma}(\gamma+i \to  (QQ')[n] + X)= \frac{{\bf \overline{\sum} }|\mathcal{M}|^2}{4\sqrt{(p_{1}+p_2)^2}|\vec{ p}_{1}|} d\Phi_j,
\end{equation}
where ${\bf \overline{\sum} }$ denotes the average of the spin and color states of initial particles and the sum of the color and spin states of all final particles, and $d\Phi_j$ represents the final $j$-body phase space element,
\begin{equation}
d\Phi_j = (2\pi)^4 \delta^4(p_1 + p_2 - \sum_{f=3}^{j+2} p_f) \prod_{f=3}^{j+2} \frac{d^3p_f}{(2\pi)^3 2p_f^0}
\end{equation}
and $\mathcal{M}$ is the total hard scattering amplitude
\begin{equation}
\mathcal{M}= { \sum_k } \mathcal{M}_k,
\end{equation}
where $k$ sums over the relevant Feynman diagrams.

There are a total of $24$ Feynman diagrams for the subprocess $\gamma + g \to (QQ')[n] + \bar{Q} + \bar{Q'}$ ($k=24$) and $4$ diagrams for $\gamma + Q \to (QQ')[n] + \bar{Q}$ ($k=4$). Additionally, for the subprocesses $\gamma + g \to (QQ)[n] + \bar{Q} + \bar{Q}$ and $\gamma + Q \to (QQ)[n] + \bar{Q}$, there are another 24 and 4 diagrams, respectively, due to the exchange of two identical quark lines within the $(QQ)[n]$-quark pair. Practically, these diagrams are equivalent to those without exchanges, as we have set the relative velocity between the two $Q$ quarks to zero, specifically $p_{31}=p_{32}=\frac{p_3}{2}$ for the production of $\Xi_{QQ}$ under the nonrelativistic approximation. A factor of $1/2!$ is included for the squared amplitude due to the identical quarks in the $(QQ)[n]$ diquark. Therefore, we only need to calculate the $24$ and $4$ diagrams for the subprocesses $\gamma + g \to (QQ)[n] + \bar{Q} + \bar{Q}$ and $\gamma + Q \to (QQ)[n] + \bar{Q}$, respectively, and then multiply by a factor of $2^2/2!$ at the cross-section level. Additionally, there is an extra factor of $1/2$ for the subprocess $\gamma + g \to (QQ)[n] + \bar{Q} + \bar{Q}$ due to the two identical open antiquarks $\bar{Q}$ in the final 3-body phase space. The amplitudes for $\gamma + g \to (QQ')[n] + \bar{Q} + \bar{Q'}$ and $\gamma + Q \to (QQ')[n] + \bar{Q'}$ can be directly derived from the Feynman diagrams. To describe the bound system of the doubly heavy baryon, we must apply spin- and color-projection operators to the amplitude of the $(QQ')[n]$-quark pair. For a detailed explanation of how to apply these projection operators and the calculation of the color factor for heavy baryon production, please refer to Ref.~\cite{Chang:2006eu}.

While the photon-light quark channel $\gamma + q \to (QQ')[n] + q + \bar{Q} + \bar{Q'}$ is theoretically possible, its contribution is expected to be negligible compared to the $\gamma + g$ and $\gamma + Q$ channels. This is due to the smaller light quark PDF, the additional suppression from phase space, and the presence of extra QCD vertices leading to reduced cross-sections. Therefore, we focus on the leading production mechanisms in this study.

To perform the calculations, we use the FeynArts package \cite{Hahn:2000kx} to generate the Feynman diagrams and amplitudes. Numerical integrations over the 2- and 3-body phase spaces are conducted using the VEGAS \cite{Lepage:1977sw} and FormCalc \cite{Hahn:1998yk} packages.

\section{Numerical results and discussions}

\subsection{Input parameters}

The matrix element $h_{\bar{\textbf{3}}}$ is linked to the Schrödinger wave function at the origin, expressed as $h_{\bar{\textbf{3}}} = |\Psi_{(QQ')}(0)|^2$ \cite{Bagan:1994dy}. According to the velocity scaling rule of NRQCD \cite{Ma:2003zk}, the color-sextuplet matrix element $h_{\textbf{6}}$ is of the same order as $h_{\bar{\textbf{3}}}$, and we use the conventional choice of ${h_\textbf{6}} = h_{\bar{\textbf{3}}}$ for our calculations. Since $h_{{\textbf{6}}/{\bar{\textbf{3}}}}$ is an overall factor, our results can be refined when more accurate values for $h_{{\textbf{6}}/{\bar{\textbf{3}}}}$ become available. The wave functions at the origin, along with the heavy quark masses, are taken as follows \cite{Baranov:1995rc, Bagan:1994dy}:

\begin{eqnarray}
&& \quad \vert \Psi_{(cc)}(0) \vert^2=0.039 ~{\rm{GeV^3}},\vert \Psi_{(bc)}(0) \vert^2=0.065 ~{\rm{GeV^3}},\nonumber \\
&&\vert \Psi_{(bb)}(0) \vert^2=0.152 ~{\rm{GeV^3}},
m_b=5.1~{\rm{GeV}}, m_c=1.8 ~{\rm{GeV}}.\nonumber
\end{eqnarray}

\begin{widetext}
    \begin{center}
    \begin{table}[htb]
    \caption{Total cross sections (in unit pb) for the production of $\Xi_{QQ'}$ at the $\sqrt{s}=1$ TeV MuIC.}
    \label{TotalCross}
    \begin{tabular}{|c|c|c|c|c|c|c|c|c|}
    \hline
    & $ ~(cc)_{6}[^{1}S_{0}]~ $ & $ ~(cc)_{\bar{3}}[^{3}S_{1}]~ $ & $ ~(bb)_{6}[^{1}S_{0}]~ $ & $ ~(bb)_{\bar{3}}[^{3}S_{1}]~ $ & $ ~(bc)_{\bar{3}}[^{1}S_{0}]~ $ & $ ~(bc)_{6}[^{1}S_{0}]~ $ & $ ~(bc)_{\bar{3}}[^{3}S_{1}]~ $ & $ ~(bc)_{6}[^{3}S_{1}]~ $\\
    \hline
    $\sigma_{\gamma g}$ & $2.19 \times 10^3$ & $2.45 \times 10^4$ & $3.31$ & $2.83 \times 10^1$ &
                        $1.55 \times 10^2$  & $1.07 \times 10^2$   &  $5.11 \times 10^2$    &$4.34 \times 10^2$ \\
    \hline
    $\sigma_{\gamma c}$ & $4.53 \times 10^3$ & $6.06 \times 10^4$ &   -   &   -   &
                        $2.82 \times 10^1$  & $1.41 \times 10^1$ & $1.33 \times 10^2$ & $6.63 \times 10^1$ \\
    \hline
    $\sigma_{\gamma b}$ &   -   &   -   &  $2.90$ & $4.02 \times 10^1$ &
                        $5.00 \times 10^2$ & $2.50 \times 10^2$ & $2.27 \times 10^3$ & $1.14 \times 10^3$ \\
    \hline
    Total  & \multicolumn{2}{c|}{$9.18 \times 10^4$} & \multicolumn{2}{c|}{$7.47\times10^1$} &\multicolumn{4}{c|}{$5.61 \times 10^3$} \\
    \hline
    \end{tabular}
    \end{table}
    \end{center}
\end{widetext}

The charm and bottom quark masses are set to $m_c = 1.80$ GeV and $m_b = 5.1$ GeV, following previous NRQCD studies. These values are slightly larger than the pole mass but are chosen to optimize the description of the heavy diquark system and the wave function at the origin $|\Psi(0)|^2$. The effective quark masses are widely used in phenomenological studies to improve agreement with experimental data.

The muon mass is taken as $m_{\mu} = 1.057 \times 10^{-1}$ GeV \cite{ParticleDataGroup:2024cfk}, and the fine-structure constant is set to $\alpha = 1/137$. The electron scattering angle cut $\theta_c$ is chosen to be 32~{\rm{mrad}}, consistent with the selections in Refs. \cite{Qiao:2003ba, Li:2009zzu}. The renormalization and factorization scales are set to the transverse mass of $\Xi_{QQ'}$, specifically $\mu_r = \mu_f = M_T$, where $M_T = \sqrt{p_T^2 + M^2}$, with $M$ being the mass of $\Xi_{QQ'}$. Here, $M = m_Q + m_{Q'}$, ensuring gauge invariance of the hard scattering amplitude. We use the nCTEQ15\_197\_79 \cite{Kovarik:2015cma} for the nucleon PDF. For the collision energies, we consider $\sqrt{s} = 1$ TeV at the MuIC.

\subsection{Basic results}

In Table \ref{TotalCross}, we present the calculated total cross sections for the production of doubly heavy baryons $\Xi_{QQ'}$ at a center-of-mass energy of $\sqrt{s} = 1$ TeV, which corresponds to a muon beam energy of $E_{\mu} = 960$ GeV and a proton beam energy of $E_{P} = 275$ GeV, at the Muon-Ion Collider. The cross sections are evaluated for production channels involving photon-gluon ($\gamma + g$), photon-charm ($\gamma + c$), and photon-bottom ($\gamma + b$) interactions. These channels represent the dominant production mechanisms in this energy regime. Additionally, by considering the contributions from each of these channels and accounting for all relevant spin-and-color configurations of the intermediate diquark state $QQ'$, we obtain
\begin{eqnarray}
&&\sigma^{\rm Total} (\Xi_{cc})=9.18 \times 10^4 ~\rm{pb}, \\
&&\sigma^{\rm Total} (\Xi_{bc})=5.61 \times 10^3 ~\rm{pb}, \\
&&\sigma^{\rm Total} (\Xi_{bb})=7.47 \times 10^1 ~\rm{pb}.
\end{eqnarray}

\begin{figure*}[htbp]
\includegraphics[width=0.4\textwidth]{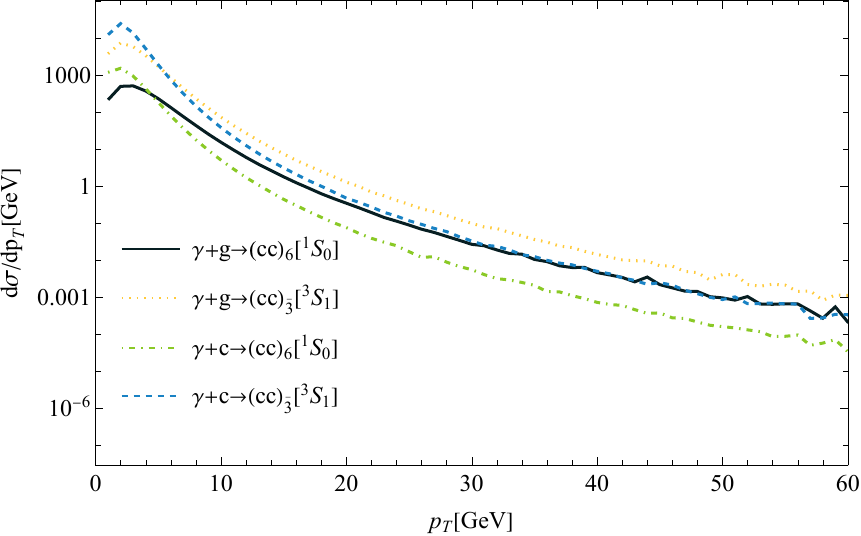}
\includegraphics[width=0.4\textwidth]{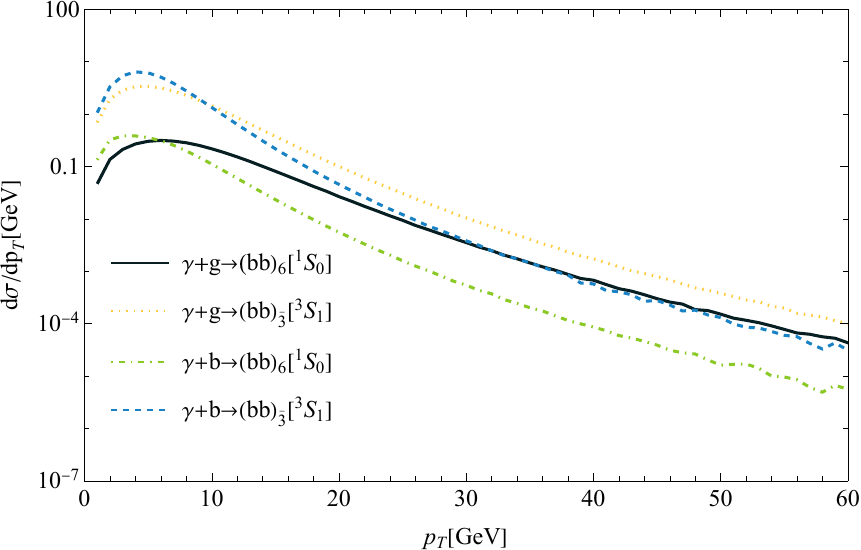}
\includegraphics[width=0.4\textwidth]{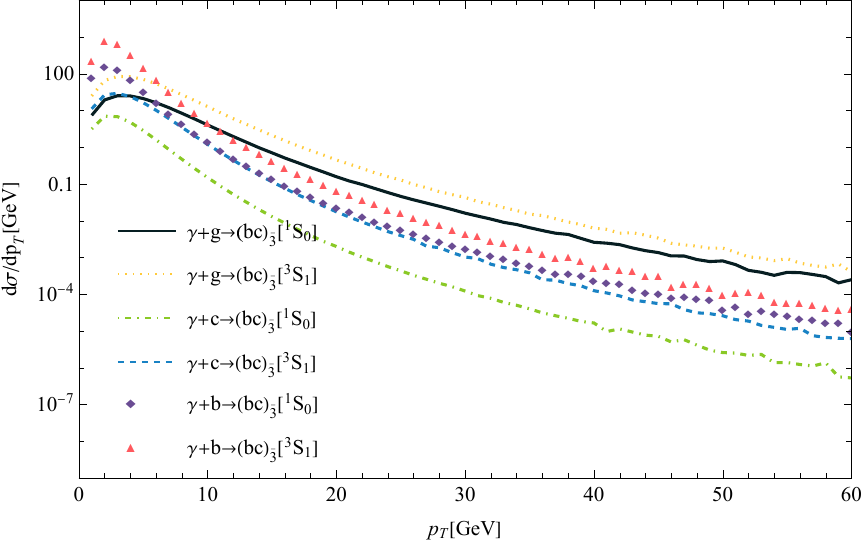}
\includegraphics[width=0.4\textwidth]{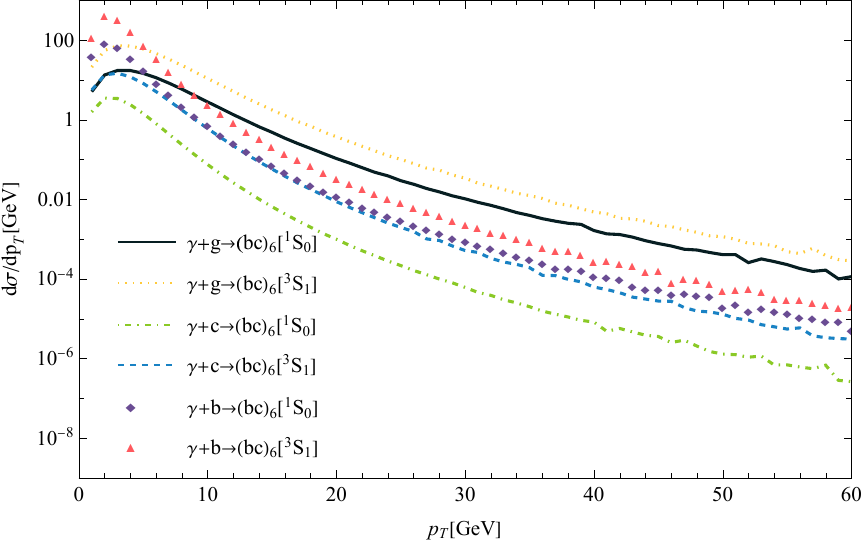}
\caption{Transverse momentum distributions for the production of $\Xi_{QQ'}$ at the $\sqrt{s}=1~{\rm TeV}$ MuIC.}
\label{figspt}
\end{figure*}
    
\begin{figure*}[htbp]
\includegraphics[width=0.4\textwidth]{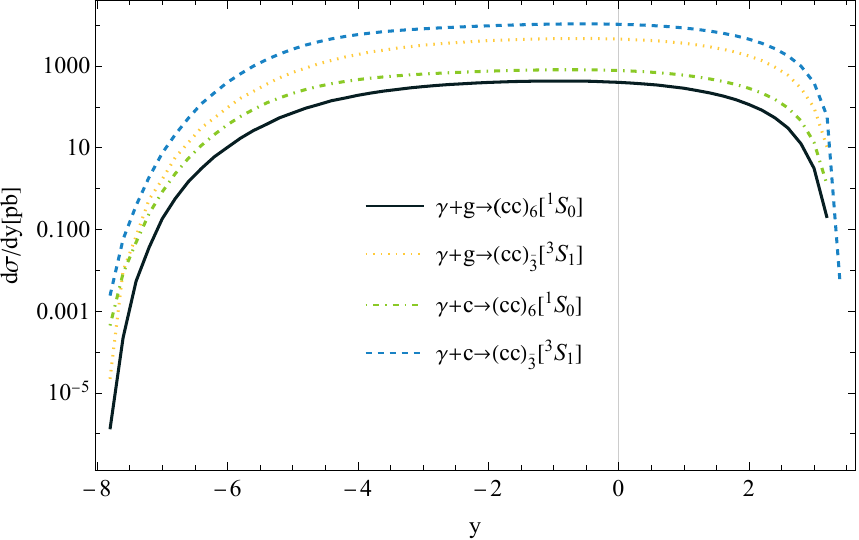}
\includegraphics[width=0.4\textwidth]{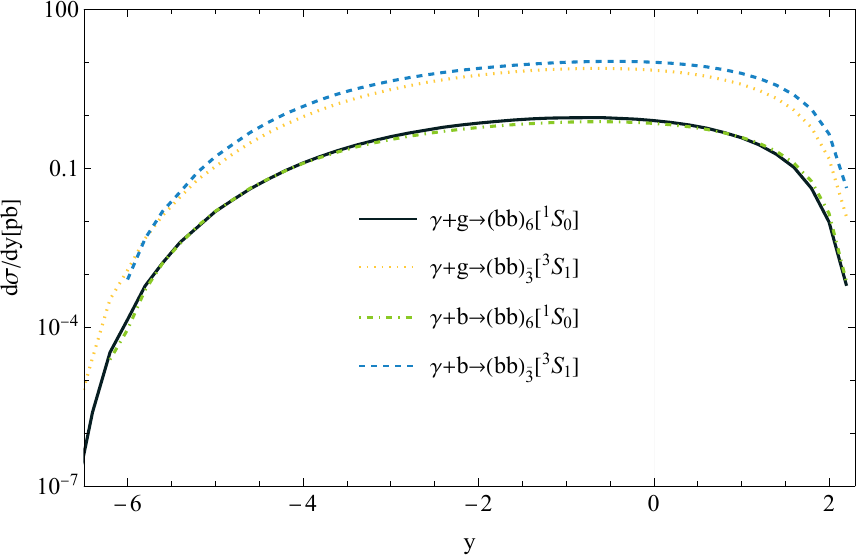}
\includegraphics[width=0.4\textwidth]{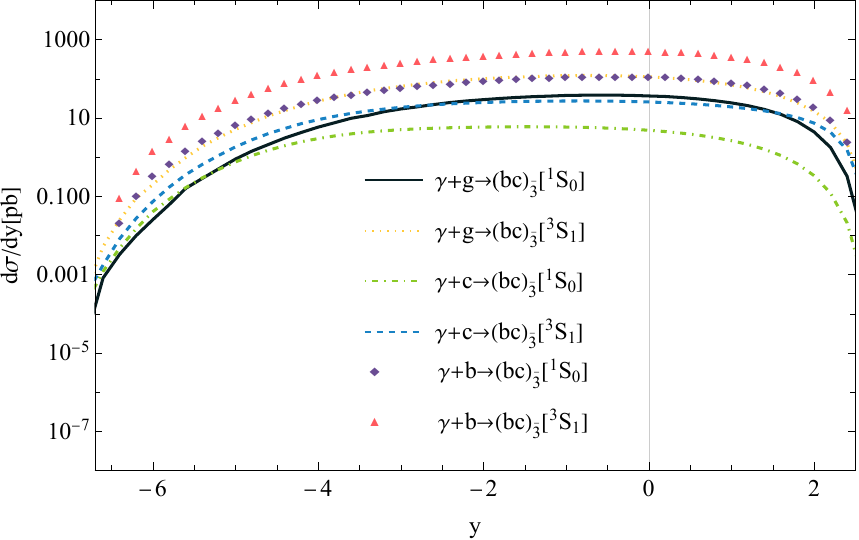}
\includegraphics[width=0.4\textwidth]{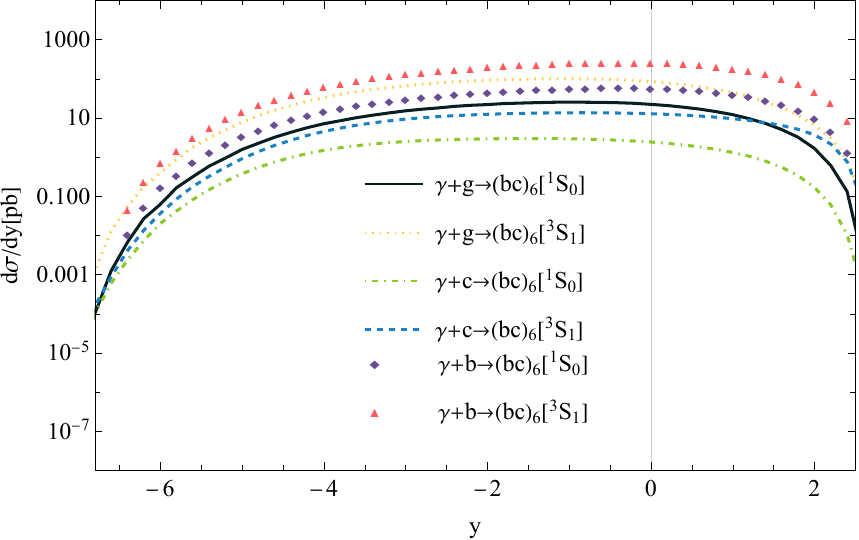}
\caption{Rapidity distributions for the production of $\Xi_{QQ'}$  at the $\sqrt{s}=1~{\rm TeV}$ MuIC.} 
\label{figsy}
\end{figure*}

\begin{figure*}[htbp]
\includegraphics[width=0.4\textwidth]{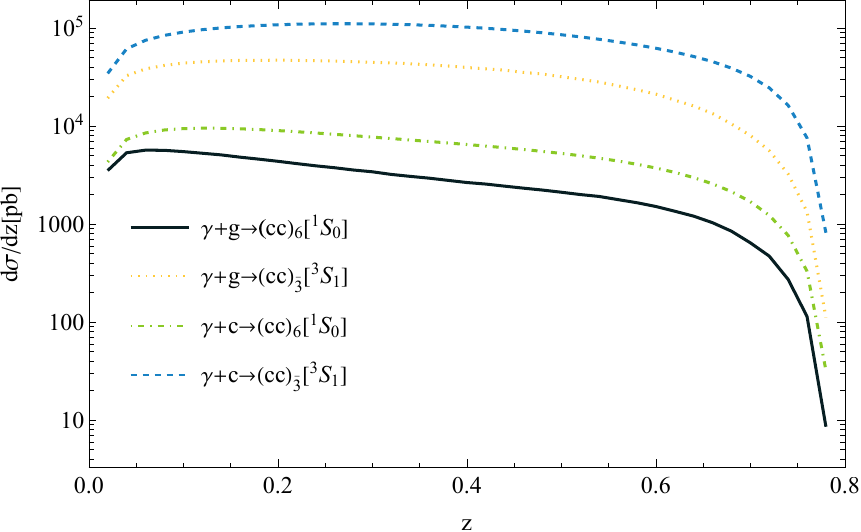}
\includegraphics[width=0.4\textwidth]{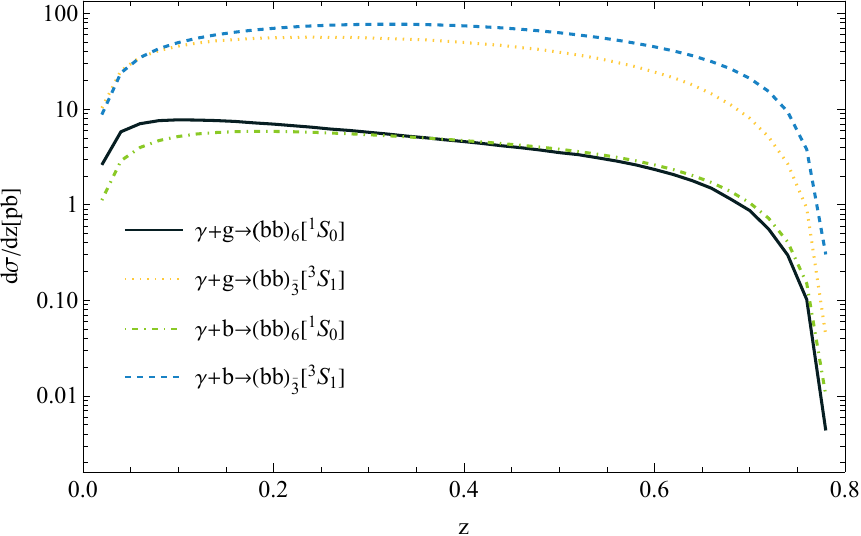}
\includegraphics[width=0.4\textwidth]{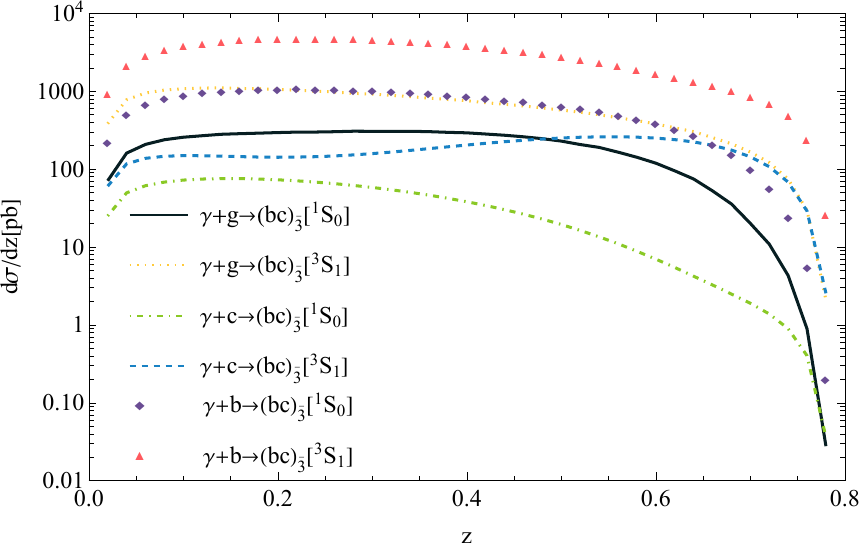}
\includegraphics[width=0.4\textwidth]{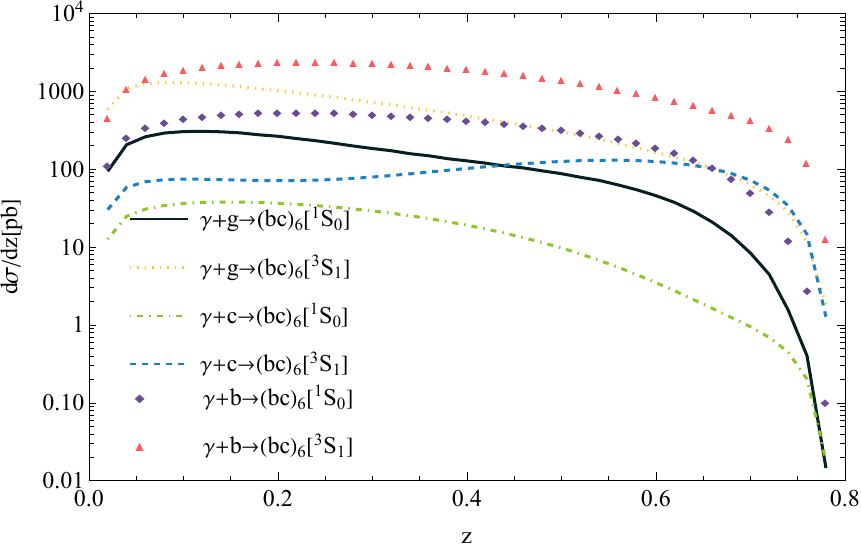}
\caption{$z$ distributions for the production of $\Xi_{QQ'}$ at the $\sqrt{s}=1$ TeV MuIC.} 
\label{figsz}
\end{figure*}

\begin{table}[htb]
\centering
\caption{Total cross sections (in units of pb) for the production of $\Xi_{QQ'}$ under different transverse momentum ($p_T$) cuts at the $\sqrt{s}=1$ TeV MuIC.}
\label{ptcuts}
\resizebox{0.4\textwidth}{!}{
\begin{tabular}{|c|c|c|c|}
\hline
& $ p_{T}\geq 1 ~\rm{GeV} $ & $ p_{T}\geq 3 ~\rm{GeV} $ & $ p_{T}\geq 5 ~\rm{GeV} $\\
\hline
$\gamma g \to (cc)_{6}[^{1}S_{0}]$ & $1.97 \times 10^3$ & $9.44 \times 10^2$ & $3.28 \times 10^2$ \\
\hline
$\gamma g \to (cc)_{\bar{3}}[^{3}S_{1}]$ & $2.06 \times 10^4$ & $7.09 \times 10^3$ & $1.89 \times 10^3$ \\
\hline
$\gamma c \to (cc)_{6}[^{1}S_{0}]$ & $3.32 \times 10^3$ & $7.89 \times 10^2$ & $1.60 \times 10^2$ \\
\hline
$\gamma c \to (cc)_{\bar{3}}[^{3}S_{1}]$ & $4.79 \times 10^4$ & $8.66 \times 10^3$ & $1.40 \times 10^3$ \\
\hline
$\gamma g \to (bb)_{6}[^{1}S_{0}]$ & $3.26$ & $2.91$ & $2.34$ \\
\hline
$\gamma g \to (bb)_{\bar{3}}[^{3}S_{1}]$ & $2.76 \times 10^1 $ & $2.28 \times 10^1$ & $1.61 \times 10^1$ \\
\hline
$\gamma b \to (bb)_{6}[^{1}S_{0}]$ & $2.77$ & $2.07$ & $1.33$ \\
\hline
$\gamma b \to (bb)_{\bar{3}}[^{3}S_{1}]$ & $3.91 \times 10^1$ & $3.05 \times 10^1$ & $1.80 \times 10^1$ \\
\hline
$\gamma g \to (bc)_{\bar{3}}[^{1}S_{0}]$ & $1.48 \times 10^2$ & $1.03 \times 10^2$ & $5.67 \times 10^1$ \\
\hline
$\gamma g \to (bc)_{6}[^{1}S_{0}]$ & $1.02 \times 10^2$ & $7.11 \times 10^1$ & $3.92 \times 10^1$ \\
\hline
$\gamma g \to (bc)_{\bar{3}}[^{3}S_{1}]$ & $4.86 \times 10^2$ & $3.35 \times 10^2$ & $1.81 \times 10^2$ \\
\hline
$\gamma g \to (bc)_{6}[^{3}S_{1}]$ & $4.13 \times 10^2$ & $2.86 \times 10^2$ & $1.55 \times 10^2$ \\
\hline
$\gamma c \to (bc)_{\bar{3}}[^{1}S_{0}]$ & $2.51 \times 10^1$ & $1.12 \times 10^1$ & $3.58$ \\
\hline
$\gamma c \to (bc)_{6}[^{1}S_{0}]$ & $1.25 \times 10^1$ & $5.61$ & $1.79$ \\
\hline
$\gamma c \to (bc)_{\bar{3}}[^{3}S_{1}]$ & $1.21 \times 10^1$ & $6.52 \times 10^1$ & $2.49 \times 10^1$ \\
\hline
$\gamma c \to (bc)_{6}[^{3}S_{1}]$ & $6.07 \times 10^1$ & $3.26 \times 10^1$ & $1.25 \times 10^1$ \\
\hline
$\gamma b \to (bc)_{\bar{3}}[^{1}S_{0}]$ & $4.23 \times 10^2$ & $1.37 \times 10^2$ & $3.49 \times 10^1$ \\
\hline
$\gamma b \to (bc)_{6}[^{1}S_{0}]$ & $2.12 \times 10^2$ & $6.84 \times 10^1$ & $1.74 \times 10^1$ \\
\hline
$\gamma b \to (bc)_{\bar{3}}[^{3}S_{1}]$ & $2.04 \times 10^3$ & $6.02 \times 10^2$ & $1.36 \times 10^2$ \\
\hline
$\gamma b \to (bc)_{6}[^{3}S_{1}]$ & $1.02 \times 10^3$ & $3.01 \times 10^2$ & $6.82 \times 10^1$ \\
\hline
\end{tabular}
}
\end{table}

\begin{table}[htbp]
\centering
\caption{Total cross sections (in units of pb) for the production of $\Xi_{QQ'}$ under different rapidity ($y$) cuts at the $\sqrt{s}=1$ TeV MuIC.}
\label{ycuts}
\resizebox{0.4\textwidth}{!}{
\begin{tabular}{|c|c|c|c|}
\hline
& $ ~~|y|\leq 1~~ $ & $ ~~|y|\leq 2~~ $ & $ ~~|y|\leq 3~~ $\\
\hline
$\gamma g \to (cc)_{6}[^{1}S_{0}]$ & $8.29 \times 10^2$ & $1.52 \times 10^3$ & $1.97 \times 10^3$ \\
\hline
$\gamma g \to (cc)_{\bar{3}}[^{3}S_{1}]$ & $8.63 \times 10^3$ & $1.56 \times 10^4$ & $2.00 \times 10^4$ \\
\hline
$\gamma c \to (cc)_{6}[^{1}S_{0}]$ & $1.46 \times 10^3$ & $2.66 \times 10^3$ & $3.46 \times 10^3$ \\
\hline
$\gamma c \to (cc)_{\bar{3}}[^{3}S_{1}]$ & $1.99 \times 10^4$ & $3.63 \times 10^4$ & $4.72 \times 10^4$ \\
\hline
$\gamma g \to (bb)_{6}[^{1}S_{0}]$ & $1.42$ & $2.37$ & $2.96$ \\
\hline
$\gamma g \to (bb)_{\bar{3}}[^{3}S_{1}]$ & $1.27 \times 10^1$ & $2.10 \times 10^1$ & $2.56 \times 10^1$ \\
\hline
$\gamma b \to (bb)_{6}[^{1}S_{0}]$ & $1.26$ & $2.09$ & $2.58$ \\
\hline
$\gamma b \to (bb)_{\bar{3}}[^{3}S_{1}]$ & $1.82 \times 10^1$ & $2.98 \times 10^1$ & $3.62 \times 10^1$ \\
\hline
$\gamma g \to (bc)_{\bar{3}}[^{1}S_{0}]$ & $6.77 \times 10^1$ & $1.15 \times 10^2$ & $1.39 \times 10^2$ \\
\hline
$\gamma g \to (bc)_{6}[^{1}S_{0}]$ & $4.05 \times 10^1$ & $7.03 \times 10^1$ & $8.98 \times 10^1$ \\
\hline
$\gamma g \to (bc)_{\bar{3}}[^{3}S_{1}]$ & $2.06 \times 10^2$ & $3.56 \times 10^2$ & $4.44 \times 10^2$ \\
\hline
$\gamma g \to (bc)_{6}[^{3}S_{1}]$ & $1.55 \times 10^2$ & $2.73 \times 10^2$ & $3.56 \times 10^2$ \\
\hline
$\gamma c \to (bc)_{\bar{3}}[^{1}S_{0}]$ & $8.92$ & $1.60 \times 10^1$ & $2.16 \times 10^1$ \\
\hline
$\gamma c \to (bc)_{6}[^{1}S_{0}]$ & $4.46$ & $7.99$ & $1.08 \times 10^1$ \\
\hline
$\gamma c \to (bc)_{\bar{3}}[^{3}S_{1}]$ & $4.85 \times 10^1$ & $8.14 \times 10^1$ & $1.11 \times 10^2$ \\
\hline
$\gamma c \to (bc)_{6}[^{3}S_{1}]$ & $2.43 \times 10^1$ & $4.37 \times 10^1$ & $5.56 \times 10^1$ \\
\hline
$\gamma b \to (bc)_{\bar{3}}[^{1}S_{0}]$ & $2.09 \times 10^2$ & $3.54 \times 10^2$ & $4.32 \times 10^2$ \\
\hline
$\gamma b \to (bc)_{6}[^{1}S_{0}]$ & $1.05 \times 10^2$ & $1.77  \times 10^2$ & $2.16 \times 10^2$ \\
\hline
$\gamma b \to (bc)_{\bar{3}}[^{3}S_{1}]$ & $9.56 \times 10^2$ & $1.62 \times 10^3$ & $1.97 \times 10^3$ \\
\hline
$\gamma b \to (bc)_{6}[^{3}S_{1}]$ & $4.78 \times 10^2$ & $8.10 \times 10^2$ & $9.86 \times 10^2$ \\
\hline
\end{tabular}
}
\end{table}

Table \ref{TotalCross} shows:
\begin{itemize}
\item For the same production channel, the $(QQ')_{\bar{\textbf{3}}} [^3S_1]$ diquark state plays a dominant role, contributing most significantly to the overall production cross section of the doubly heavy baryon $\Xi_{QQ'}$. This is particularly evident across all production mechanisms considered. In the specific cases of $\Xi_{cc}$ and $\Xi_{bb}$ production, the color-sextuplet diquark configurations, namely $(cc)_{\textbf{6}} [^1S_0]$ and $(bb)_{\textbf{6}} [^1S_0]$, contribute approximately $7\%-8\%$ of the total cross section for the same production channel. Although smaller than the contribution of the color-antitriplet configuration, this remains a non-negligible fraction. For the production of $\Xi_{bc}$, the situation is somewhat more complex, as contributions from several other diquark states become increasingly important. Notably, the cross section associated with the $(bc)_{\textbf{6}} [^3S_1]$ diquark state is comparable in magnitude to that of $(bc)_{\bar{\textbf{3}}} [^3S_1]$, highlighting the complexity of the production dynamics in this case. Therefore, a comprehensive and detailed discussion of all possible diquark configurations is crucial for achieving accurate and reliable predictions concerning the production cross sections of doubly heavy baryons in various channels.

\item In addition to the widely studied $\gamma + g$ channel, the extrinsic heavy quark mechanism, represented by the $\gamma + Q$ channel, also makes a substantial contribution to the total production cross section of doubly heavy baryons. This contribution is particularly significant in specific production scenarios. For example, in the case of $\Xi_{cc}$ production, the cross section from the $\gamma + c$ channel can exceed that of the $\gamma + g$ channel under certain kinematic conditions. This observed dominance of the $\gamma + Q$ channel, as will be elaborated upon in subsequent sections, stems primarily from its enhanced contribution in the low transverse momentum ($p_T$) region, where the extrinsic heavy quark mechanisms become more prevalent.

\end{itemize}

In Fig.\ref{figspt}, we show the transverse momentum ($p_T$) distributions for the photoproduction of ${\Xi_{QQ'}}$. Each channel exhibits a $p_T$ peak around ${\cal O}(1)~{\rm GeV}$, followed by a logarithmic decline. The $p_T$ distributions for the $\gamma + c$ and $\gamma + b$ channels decrease more rapidly than those of $\gamma + g$ in the high $p_T$ region. For the same diquark configuration, the $\gamma + Q$ channel dominates over the $\gamma + g$ channel in the low $p_T$ region, explaining the relatively large cross section observed for the $\gamma + Q$ channel in Table \ref{TotalCross}. Additionally, we observe that the $p_T$ distributions of $\Xi_{bb}$ decrease at a slower rate compared to $\Xi_{bc}$ and $\Xi_{cc}$ as $p_T$ increases, suggesting that $\Xi_{bb}$ events could become comparable to $\Xi_{cc}$ and $\Xi_{bc}$ in the high $p_T$ region, despite the total cross section of $\Xi_{bb}$ being considerably smaller than those of $\Xi_{cc}$ and $\Xi_{bc}$.

In Fig.\ref{figsy}, we present the rapidity ($y$) distributions for the photoproduction of ${\Xi_{QQ'}}$. The pronounced asymmetry in the rapidity distributions for $\Xi_{QQ'}$ clearly indicates that the dominant production occurs in the region where $y<0$. The $z$-axis in our setup is aligned with the electron beam, and the observation of $y<0$ signifies that the parton $i$ originating from the proton carries more energy than the incoming photon, resulting in the majority of $\Xi_{QQ'}$ events being produced in the direction of the proton beam.

In Tables \ref{ptcuts} and \ref{ycuts}, we present the calculated cross sections for the production of doubly heavy baryons under a range of transverse momentum ($p_T$) and rapidity ($y$) cuts. Table \ref{ptcuts} shows that the cross sections for $\Xi_{bc}$ and $\Xi_{cc}$ exhibit greater sensitivity to varying $p_T$ cuts compared to those of $\Xi_{bb}$, a trend consistent with the behavior observed in Fig.\ref{figspt}. Furthermore, as indicated by Table \ref{ptcuts}, even with relatively large $p_T$ cuts, a substantial production rate for $\Xi_{bc}$ and $\Xi_{cc}$ can still be achieved.

Finally, we illustrate the differential cross sections $d\sigma / dz$ for various diquark configurations and production channels in Fig.\ref*{figsz}, where $z = \frac{p_{\Xi} \cdot p_P}{p_{\gamma} \cdot p_P}$. In this expression, $p_{\gamma}$, $p_P$, and $p_{\Xi}$ denote the four-momenta of the photon, proton, and $\Xi_{QQ'}$, respectively.

\subsection{Theoretical uncertainties}

\begin{table}[htbp]
\centering
\caption{The variations in the total cross sections (in units of pb) for the production of $\Xi_{cc}$ and $\Xi_{bc}$ at the $\sqrt{s}=1$ TeV MuIC are analyzed with $m_c = 1.80 \pm 0.10 , \rm{GeV}$. The mass $m_b$ is held constant at 5.10 GeV while examining the uncertainty stemming from $m_c$.}
\label{mc}
\resizebox{0.4\textwidth}{!}{
\begin{tabular}{|c|c|c|c|}
\hline
& $ m_{c}=1.7 \rm{GeV} $ & $ m_{c}=1.8 \rm{GeV} $ & $ m_{c}=1.9 \rm{GeV} $\\
\hline
$\gamma g \to (cc)_{6}[^{1}S_{0}]$ & $3.14 \times 10^3$ & $2.19 \times 10^3$ & $1.56 \times 10^3$ \\
\hline
$\gamma g \to (cc)_{\bar{3}}[^{3}S_{1}]$ & $3.50 \times 10^4$ & $2.45 \times 10^4$ & $1.75 \times 10^4$ \\
\hline
$\gamma c \to (cc)_{6}[^{1}S_{0}]$ & $6.00 \times 10^3$ & $4.53  \times 10^3$ & $3.46 \times 10^3$ \\
\hline
$\gamma c \to (cc)_{\bar{3}}[^{3}S_{1}]$ & $8.02 \times 10^4$ & $6.06 \times 10^4$ & $4.63 \times 10^4$ \\
\hline
$\gamma g \to (bc)_{\bar{3}}[^{1}S_{0}]$ & $1.85 \times 10^2$ & $1.55 \times 10^2$ & $1.31 \times 10^2$ \\
\hline
$\gamma g \to (bc)_{6}[^{1}S_{0}]$ & $1.29  \times 10^2$ & $1.07 \times 10^2$ & $8.96 \times 10$ \\
\hline
$\gamma g \to (bc)_{\bar{3}}[^{3}S_{1}]$ & $6.06 \times 10^2$ & $5.11 \times 10^2$ & $4.35 \times 10^2$ \\
\hline
$\gamma g \to (bc)_{6}[^{3}S_{1}]$ & $5.17 \times 10^2$ & $4.34 \times 10^2$ & $3.69 \times 10^2$ \\
\hline
$\gamma c \to (bc)_{\bar{3}}[^{1}S_{0}]$ & $2.98 \times 10^1$ & $2.82  \times 10^1$ & $2.67 \times 10^1$ \\
\hline
$\gamma c \to (bc)_{6}[^{1}S_{0}]$ & $1.49 \times 10^1$ & $1.41 \times 10^1$ & $1.33 \times 10^1$ \\
\hline
$\gamma c \to (bc)_{\bar{3}}[^{3}S_{1}]$ & $1.41 \times 10^2$ & $1.33 \times 10^2$ & $1.25 \times 10^2$ \\
\hline
$\gamma c \to (bc)_{6}[^{3}S_{1}]$ & $7.04 \times 10^1$ & $6.63 \times 10^1$ & $6.24 \times 10^1$ \\
\hline
$\gamma b \to (bc)_{\bar{3}}[^{1}S_{0}]$ & $6.59 \times 10^2$ & $5.00 \times 10^2$ & $3.85 \times 10^2$ \\
\hline
$\gamma b \to (bc)_{6}[^{1}S_{0}]$ & $3.29 \times 10^2$ & $2.50 \times 10^2$ & $1.93 \times 10^2$ \\
\hline
$\gamma b \to (bc)_{\bar{3}}[^{3}S_{1}]$ & $2.96 \times 10^3$ & $2.27 \times 10^3$ & $1.77 \times 10^3$ \\
\hline
$\gamma b \to (bc)_{6}[^{3}S_{1}]$ & $1.48 \times 10^3$ & $1.14 \times 10^3$ & $8.86 \times 10^2$ \\
\hline
\end{tabular}
}
\end{table}

\begin{table}[htbp]
\centering
\caption{The variations in the total cross sections (in units of pb) for the production of $\Xi_{bb}$ and $\Xi_{bc}$ at the $\sqrt{s}=1$ TeV MuIC are examined with $m_b = 5.10 \pm 0.20 , \rm{GeV}$. The charm quark mass $m_c$ is fixed at 1.8 GeV while assessing the uncertainty associated with $m_b$.}
\label{mb}
\resizebox{0.4\textwidth}{!}{
\begin{tabular}{|c|c|c|c|}
\hline
& $ m_{b}=4.9 \rm{GeV} $ & $ m_{b}=5.1 \rm{GeV} $ & $ m_{b}=5.3 \rm{GeV} $\\
\hline
$\gamma g \to (bb)_{6}[^{1}S_{0}]$ & $4.33$ & $3.31$ & $2.54$ \\
\hline
$\gamma g \to (bb)_{\bar{3}}[^{3}S_{1}]$ & $3.70 \times 10^1$ & $2.83 \times 10^1$ & $2.18 \times 10^1$ \\
\hline
$\gamma b \to (bb)_{6}[^{1}S_{0}]$ & $3.36$ & $2.90$ & $2.49$ \\
\hline
$\gamma b \to (bb)_{\bar{3}}[^{3}S_{1}]$ & $4.65 \times 10^1$ & $4.02 \times 10^1$ & $3.45 \times 10^1$ \\
\hline
$\gamma g \to (bc)_{\bar{3}}[^{1}S_{0}]$ & $1.79 \times 10^2$ & $1.55 \times 10^2$ & $1.35 \times 10^2$ \\
\hline
$\gamma g \to (bc)_{6}[^{1}S_{0}]$ & $1.22 \times 10^2$ & $1.07 \times 10^2$ & $9.41 \times 10^2$ \\
\hline
$\gamma g \to (bc)_{\bar{3}}[^{3}S_{1}]$ & $5.90 \times 10^2$ & $5.11 \times 10^2$ & $4.46 \times 10^1$ \\
\hline
$\gamma g \to (bc)_{6}[^{3}S_{1}]$ & $5.00 \times 10^2$ & $4.34 \times 10^2$ & $3.80 \times 10^2$ \\
\hline
$\gamma c \to (bc)_{\bar{3}}[^{1}S_{0}]$ & $3.33 \times 10^1$ & $2.82 \times 10^1$ & $2.40 \times 10^1$ \\
\hline
$\gamma c \to (bc)_{6}[^{1}S_{0}]$ & $1.67 \times 10^1$ & $1.41 \times 10^1$ & $1.20 \times 10^1$ \\
\hline
$\gamma c \to (bc)_{\bar{3}}[^{3}S_{1}]$ & $1.56 \times 10^2$ & $1.33 \times 10^2$ & $1.13 \times 10^2$ \\
\hline
$\gamma c \to (bc)_{6}[^{3}S_{1}]$ & $7.80 \times 10^1$ & $6.63 \times 10^1$ & $5.66 \times 10^1$ \\
\hline
$\gamma b \to (bc)_{\bar{3}}[^{1}S_{0}]$ & $4.50 \times 10^2$ & $5.00 \times 10^2$ & $5.42 \times 10^2$ \\
\hline
$\gamma b \to (bc)_{6}[^{1}S_{0}]$ & $2.25 \times 10^2$ & $2.50 \times 10^2$ & $2.71 \times 10^2$ \\
\hline
$\gamma b \to (bc)_{\bar{3}}[^{3}S_{1}]$ & $2.06 \times 10^3$ & $2.27 \times 10^3$ & $2.44 \times 10^3$ \\
\hline
$\gamma b \to (bc)_{6}[^{3}S_{1}]$ & $1.03 \times 10^3$ & $1.14 \times 10^3$ & $1.22 \times 10^3$ \\
\hline
\end{tabular}
}
\end{table}

\begin{table}[htbp]
\centering
\caption{The variations in the total cross sections (in units of pb) for the photoproduction of $\Xi_{QQ'}$ at the $\sqrt{s}=1$ TeV MuIC are analyzed with the renormalization/factorization scale set at $\mu=0.75M_T$ and $\mu=1.25M_T$.}
\label{scaleu}
\resizebox{0.4\textwidth}{!}{
\begin{tabular}{|c|c|c|c|}
\hline
& $ \mu = 0.75 M_{T} $ & $ \mu = M_{T} $ & $ \mu = 1.25 M_{T} $\\
\hline
$\gamma g \to (cc)_{6}[^{1}S_{0}]$ & $2.52 \times 10^3$ & $2.19 \times 10^3$ & $1.98 \times 10^3$ \\
\hline
$\gamma g \to (cc)_{\bar{3}}[^{3}S_{1}]$ & $2.79 \times 10^4$ & $2.45 \times 10^4$ & $2.22 \times 10^4$ \\
\hline
$\gamma c \to (cc)_{6}[^{1}S_{0}]$ & $4.94 \times 10^3$ & $4.53 \times 10^3$ & $4.20 \times 10^3$ \\
\hline
$\gamma c \to (cc)_{\bar{3}}[^{3}S_{1}]$ & $6.60 \times 10^4$ & $6.06 \times 10^4$ & $5.62 \times 10^4$ \\
\hline
$\gamma g \to (bb)_{6}[^{1}S_{0}]$ & $3.87$ & $3.31$ & $2.94$ \\
\hline
$\gamma g \to (bb)_{\bar{3}}[^{3}S_{1}]$ & $3.31 \times 10^1$ & $2.83 \times 10^1$ & $2.52 \times 10^1$ \\
\hline
$\gamma b \to (bb)_{6}[^{1}S_{0}]$ & $2.74$ & $2.90$ & $2.94$ \\
\hline
$\gamma b \to (bb)_{\bar{3}}[^{3}S_{1}]$ & $3.79 \times 10^1$ & $4.02 \times 10^1$ & $4.08 \times 10^1$ \\
\hline
$\gamma g \to (bc)_{\bar{3}}[^{1}S_{0}]$ & $1.81 \times 10^2$ & $1.55 \times 10^2$ & $1.38 \times 10^2$ \\
\hline
$\gamma g \to (bc)_{6}[^{1}S_{0}]$ & $1.24 \times 10^2$ & $1.07 \times 10^2$ & $9.56 \times 10^1$ \\
\hline
$\gamma g \to (bc)_{\bar{3}}[^{3}S_{1}]$ & $5.93 \times 10^2$ & $5.11 \times 10^2$ & $4.57 \times 10^2$ \\
\hline
$\gamma g \to (bc)_{6}[^{3}S_{1}]$ & $5.03 \times 10^2$ & $4.34 \times 10^2$ & $3.89 \times 10^2$ \\
\hline
$\gamma c \to (bc)_{\bar{3}}[^{1}S_{0}]$ & $3.10 \times 10^1$ & $2.82 \times 10^1$ & $2.62 \times 10^1$ \\
\hline
$\gamma c \to (bc)_{6}[^{1}S_{0}]$ & $1.55 \times 10^1$ & $1.41 \times 10^1$ & $1.31 \times 10^1$ \\
\hline
$\gamma c \to (bc)_{\bar{3}}[^{3}S_{1}]$ & $1.46 \times 10^2$ & $1.33 \times 10^2$ & $1.23 \times 10^2$ \\
\hline
$\gamma c \to (bc)_{6}[^{3}S_{1}]$ & $7.28 \times 10^1$ & $6.63 \times 10^1$ & $6.16 \times 10^1$ \\
\hline
$\gamma b \to (bc)_{\bar{3}}[^{1}S_{0}]$ & $3.59 \times 10^2$ & $5.00 \times 10^2$ & $5.46 \times 10^2$ \\
\hline
$\gamma b \to (bc)_{6}[^{1}S_{0}]$ & $1.80 \times 10^2$ & $2.50 \times 10^2$ & $2.73 \times 10^2$ \\
\hline
$\gamma b \to (bc)_{\bar{3}}[^{3}S_{1}]$ & $1.64 \times 10^3$ & $2.27 \times 10^3$ & $2.48 \times 10^3$ \\
\hline
$\gamma b \to (bc)_{6}[^{3}S_{1}]$ & $8.18 \times 10^2$ & $1.14 \times 10^3$ & $1.24 \times 10^3$ \\
\hline
\end{tabular}
}
\end{table}

\begin{table}[htbp]
\centering
\caption{The variations in the total cross sections (in units of pb) for the photoproduction of $\Xi_{QQ'}$ at the $\sqrt{s}=1$ TeV MuIC are analyzed by setting the electron scattering angle cut at $\theta_c=64$ and $16$ mrad.}
\label{thetac}
\resizebox{0.4\textwidth}{!}{
\begin{tabular}{|c|c|c|c|}
\hline
& $ \theta_{c} = 16 ~\rm{mrad} $ & $ \theta_{c} = 32 ~\rm{mrad} $ & $ \theta_{c} = 64 ~\rm{mrad} $\\
\hline
$\gamma g \to (cc)_{6}[^{1}S_{0}]$ & $1.98 \times 10^3$ & $2.19 \times 10^3$ & $2.40 \times 10^3$ \\
\hline
$\gamma g \to (cc)_{\bar{3}}[^{3}S_{1}]$ & $2.22 \times 10^4$ & $2.45 \times 10^4$ & $2.68 \times 10^4$ \\
\hline
$\gamma c \to (cc)_{6}[^{1}S_{0}]$ & $4.12 \times 10^3$ & $4.53 \times 10^3$ & $4.94 \times 10^3$ \\
\hline
$\gamma c \to (cc)_{\bar{3}}[^{3}S_{1}]$ & $5.52 \times 10^4$ & $6.06 \times 10^4$ & $6.60 \times 10^4$ \\
\hline
$\gamma g \to (bb)_{6}[^{1}S_{0}]$ & $3.05$ & $3.31$ & $3.56$ \\
\hline
$\gamma g \to (bb)_{\bar{3}}[^{3}S_{1}]$ & $2.52 \times 10^1$ & $2.83 \times 10^1$ & $3.13 \times 10^1$ \\
\hline
$\gamma b \to (bb)_{6}[^{1}S_{0}]$ & $2.59$ & $2.90$ & $3.21$ \\
\hline
$\gamma b \to (bb)_{\bar{3}}[^{3}S_{1}]$ & $3.59 \times 10^1$ & $4.02 \times 10^1$ & $4.45 \times 10^1$ \\
\hline
$\gamma g \to (bc)_{\bar{3}}[^{1}S_{0}]$ & $1.39 \times 10^2$ & $1.55 \times 10^2$ & $1.71 \times 10^2$ \\
\hline
$\gamma g \to (bc)_{6}[^{1}S_{0}]$ & $9.59 \times 10^1$ & $1.07 \times 10^2$ & $1.18 \times 10^2$ \\
\hline
$\gamma g \to (bc)_{\bar{3}}[^{3}S_{1}]$ & $4.59 \times 10^2$ & $5.11 \times 10^2$ & $5.64 \times 10^2$ \\
\hline
$\gamma g \to (bc)_{6}[^{3}S_{1}]$ & $3.90 \times 10^2$ & $4.34 \times 10^2$ & $4.79 \times 10^2$ \\
\hline
$\gamma c \to (bc)_{\bar{3}}[^{1}S_{0}]$ & $2.55 \times 10^1$ & $2.82 \times 10^1$ & $3.09 \times 10^1$ \\
\hline
$\gamma c \to (bc)_{6}[^{1}S_{0}]$ & $1.28 \times 10^1$ & $1.41 \times 10^1$ & $1.55 \times 10^1$ \\
\hline
$\gamma c \to (bc)_{\bar{3}}[^{3}S_{1}]$ & $1.20 \times 10^2$ & $1.33 \times 10^2$ & $1.45 \times 10^2$ \\
\hline
$\gamma c \to (bc)_{6}[^{3}S_{1}]$ & $6.00 \times 10^1$ & $6.63 \times 10^1$ & $7.26 \times 10^1$ \\
\hline
$\gamma b \to (bc)_{\bar{3}}[^{1}S_{0}]$ & $4.49 \times 10^2$ & $5.00 \times 10^2$ & $5.50 \times 10^2$ \\
\hline
$\gamma b \to (bc)_{6}[^{1}S_{0}]$ & $2.25 \times 10^2$ & $2.50 \times 10^2$ & $2.75 \times 10^2$ \\
\hline
$\gamma b \to (bc)_{\bar{3}}[^{3}S_{1}]$ & $2.04 \times 10^3$ & $2.27 \times 10^3$ & $2.50 \times 10^3$ \\
\hline
$\gamma b \to (bc)_{6}[^{3}S_{1}]$ & $1.02 \times 10^3$ & $1.14 \times 10^3$ & $1.25 \times 10^3$ \\
\hline
\end{tabular}
}
\end{table}

The non-perturbative matrix elements act as overall parameters, and their uncertainties can be effectively minimized when their precise values are known. In this subsection, we discuss the uncertainties arising from the $c$-quark mass, the $b$-quark mass, the renormalization (factorization) scale, and the scattering angle cut $\theta_c$. For clarity, when examining the uncertainty of one parameter, we will keep the other parameters at their central values.

We present the cross sections for $\Xi_{QQ'}$ at the $\sqrt{s}=1$ TeV MuIC for $m_c = 1.80 \pm 0.10 $ GeV in Table \ref*{mc} and $m_b = 5.10 \pm 0.20$ GeV in Table \ref*{mb}. Both tables show that the cross sections decrease as the masses of the charm and bottom quarks increase, except for the channel $\gamma + b \to \Xi_{bc}$, which shows an increase with a rising bottom quark mass. Additionally, the data indicate that the uncertainty in the mass of $\Xi_{bc}$ is particularly sensitive to variations in the charm quark mass. By combining these two uncertainties in quadrature and summing over all diquark configurations and production channels, we obtain
\begin{eqnarray}
&&\sigma^{\rm Total} (\Xi_{cc})=9.18^{+3.22}_{-2.28} \times 10^4 ~\rm{pb},\nonumber \\
&&\sigma^{\rm Total} (\Xi_{bc})=5.6^{+0.7}_{-0.5} \times 10^3 ~\rm{pb},\nonumber \\
&&\sigma^{\rm Total} (\Xi_{bb})=7.5^{+1.6}_{-1.4} \times 10^1 ~\rm{pb}.\nonumber
\end{eqnarray}

Setting the renormalization scale is crucial for fixed-order pQCD predictions \cite{Wu:2013ei}. To investigate scale uncertainty, we set the factorization scale equal to the renormalization scale, $\mu_f=\mu_r=\mu$. Besides the choice of $\mu=M_T$, we examine two other scales: $\mu=0.75M_T$ and $\mu=1.25M_T$. Table \ref{scaleu} outlines the scale uncertainties for each diquark configuration and production channel. When the scale changes from $M_T$ to $0.75M_T$, the uncertainties in the total cross sections are roughly $10\%$, $17\%$, and $4\%$ for $\Xi_{cc}$, $\Xi_{bc}$, and $\Xi_{bb}$, respectively. Conversely, varying the scale from $M_T$ to $1.25M_T$ results in uncertainties of approximately $8\%$, $4\%$, and $4\%$ for $\Xi_{cc}$, $\Xi_{bc}$, and $\Xi_{bb}$.

Lastly, we discuss the uncertainties arising from the scattering angle cut $\theta_c$. For this analysis, we set $\theta_c=16$ and $64$ mrad. The results are presented in Table \ref{thetac}, indicating that the uncertainties due to $\theta_c$ are approximately $9\%$ for $\Xi_{cc}$, $10\%$ for $\Xi_{bc}$, and $10\%$ for $\Xi_{bb}$. The relatively small uncertainty from $\theta_c$ suggests that the photon density function is a suitable choice for our calculations.

\section{Summary}

This paper examines the production of doubly heavy baryons at the $\sqrt{s} = 1 $ TeV MuIC through the channels $\gamma + g \to \Xi_{QQ'} + \bar{Q} + \bar{Q'}$ and $\gamma + Q \to \Xi_{QQ'} + \bar{Q'}$ within the NRQCD framework. Our findings indicate that the extrinsic heavy quark mechanism via $\gamma + Q \to \Xi_{QQ'} + \bar{Q'}$ results in a significantly higher production rate compared to the $\gamma + g \to \Xi_{QQ'} + \bar{Q} + \bar{Q'}$ channel, even with the suppression from heavy quark parton distribution functions (PDFs). There are four spin-and-color diquark configurations for producing doubly heavy baryons: $(QQ')_{\bar{\textbf{3}}/{\textbf{6}}} [^1S_0]$ and $(QQ')_{\bar{\textbf{3}}/{\textbf{6}}} [^3S_1]$. Notably, the $(QQ')_{\bar{\textbf{3}}} [^3S_1]$ diquark state contributes the most to $\Xi_{QQ'}$ production, while other diquark states also significantly impact this production, underscoring the importance of discussing all configurations for accurate predictions.

We present the cross sections and the associated uncertainties arising from various choices of heavy-quark mass, renormalization/factorization scales, and $\theta_c$. By selecting $m_c=1.80\pm0.10$ GeV and $m_b=5.1\pm0.20$ GeV, we expect to generate $(3.67^{+1.29}_{-0.91})\times 10^{9}$ $\Xi_{cc}$, $(2.24^{+0.28}_{-0.20})\times 10^{8}$ $\Xi_{bc}$, and $(3.00^{+0.64}_{-0.56})\times 10^{6}$ $\Xi_{bb}$ events in one operational year at the MuIC, under conditions of $\sqrt{s}=1$ TeV and ${\cal L}\simeq 40$ fb$^{-1}$, where all configurations and production channels have been accounted for. With these production rates, the MuIC will serve as an excellent platform for exploring the properties of doubly heavy baryons $\Xi_{QQ'}$.

\hspace{2cm}

{\bf Acknowledgement:} This work was supported in part by the Natural Science Foundation of China under Grant No. 12347101, by the Fundamental Research Funds for the Central Universities under Grant No. 2024CDJXY022, by the Chongqing Natural Science Foundation under Grant No. CSTB2022NSCQ-MSX0415. The work of H.Y Bi was supported by China Postdoctoral Science Foundation under Grant No. 2022TQ0012 and No. 2023M730097.

\end{document}